\documentclass[aps,prl,twocolumn,floats,showpacs]{revtex4}
\usepackage{epsfig}
\usepackage{amsmath}
\begin{document}
\title{
Dynamical symmetries in Kondo tunneling through complex quantum
dots }
\author{
T. Kuzmenko$^1$, K. Kikoin$^1$ and Y. Avishai$^{1,2}$\\
} \affiliation {$^1$Department of Physics and $^2$Ilse Katz
Center, Ben-Gurion University, Beer-Sheva, Israel }
%\date{\today}
\begin{abstract}
Kondo tunneling reveals hidden $SO(n)$ dynamical symmetries of
evenly occupied quantum dots.  As is exemplified for an
experimentally realizable triple quantum dot in parallel geometry,
the possible values $n=3,4,5,7$ can be easily tuned by gate
voltages. Following construction of the corresponding $o_n$
algebras,  scaling equations are derived and Kondo temperatures
are calculated. The symmetry group for a magnetic field induced
anisotropic Kondo tunneling is $SU(2)$ or $SO(4)$.
\end{abstract}
 \maketitle While theoretical
predictions of the Kondo effect in tunneling through quantum dots
(QD) under strong Coulomb blockade conditions \cite{Glaz} have
been confirmed \cite{Gogo}, it should be born in mind that
representing a {\it real} nanoobject by a {\it single} localized
spin S=1/2 is inadequate. Ubiquitous low-lying spin excitations in
few-electron systems cannot be ignored. Even in "classical" planar
QDs formed in GaAs/GaAlAs heterostructures the Kondo physics is
much richer than that employed in analyzing the seminal
experiments \cite{Gogo}.

The purpose of the present work is to demonstrate that if
low-lying spin excitations are properly incorporated, the exchange
Hamiltonian of quantum dots with even occupation ${\cal N}$
unveils an unusual dynamical $SO(n)$ symmetry, and to suggest
experiments for its elucidation. Analysis of relatively simple QD
systems indicates the possible emergence of higher symmetries. For
example, Kondo tunneling may be induced by external magnetic field
in planar QD \cite{Misha}, since occurrence of low-lying triplet
exciton above singlet ground state leads to an $SO(4)$ symmetry.
Due to Zeeman splitting, it is reduced to $SU(2)$, leading to the
Kondo effect in strong magnetic field. Similar scenario may be
realized in vertical QDs \cite{Eto} where now the Larmor (instead
of the Zeeman) effect comes into play. Another example is a double
quantum dot with ${\cal N}=2$ which is a spin analog of a hydrogen
molecule ${\rm H_2}$. Here, the low lying singlet/triplet manifold
possesses the symmetry $SO(4)$ of a "spin rotator"
\cite{KA01,KA02}.

The central (and fundamental) question is then: {\it Is the
physics of Kondo tunneling through complex quantum dots intimately
related with hidden $SO(n)$ symmetries?} The answer given below is
affirmative. Moreover, these symmetries can be experimentally
realized and the specific value of $n$ can be controlled by gate
voltage and/or tunneling strength. To be concrete, the analysis is
carried out below for a triple quantum dot (TQD) in a parallel
geometry with ${\cal N}=4$ as a neutral ground state (see Fig. 1).
It is shown to exhibit an $SO(n)$ symmetry, and the relations of
tunneling strengths $V_{l,r}$ and gate voltages $V_{gl},V_{gr}$
with the possible values $n=3,4,5,7$ and the corresponding Kondo
temperatures are explicitly demonstrated.
 This example is simple enough to allow
the construction of the corresponding $o_n$ algebras and solving
the poor-man scaling equations for obtaining the Kondo
temperatures. At the same time, it paves the way for treating more
general QD structures with arbitrary scheme of low lying spin
excitations.

Initially, the TQD in Fig. 1 is treated within an Anderson-type
model with bare level operators $d_{\sigma i}$, energies
$\varepsilon_{i}$, charging energies $Q_{i}$ and gate voltages
$V_{gi}$ with $i=l,c,r$ for left, center and right dots. The
figure also defines inter-dot hopping ($W_{a}$) and tunneling
matrix elements ($V_{a}$) where the notation $a=l,r$ and ${\bar
a}=r,l$ is used ubiquitously hereafter. It is useful to shift the
energies as
 $\epsilon_{i} = \varepsilon_{i}
-V_{gi}$ which can be experimentally manipulated. Setting the
Fermi energy in the leads to be $\varepsilon_{F}=0$, the pertinent
``Kondo limit'' is determined as, $0 > \epsilon_{a} \gg
\epsilon_{c}$ and $0 < \epsilon_{a}+Q_{a} \ll \epsilon_{c}+Q_{c}$
\cite{KA01}. The capacitive interaction between the three dots is
tuned in such a way that, in the absence of inter-dot hopping, the
neutral ground state has the occupation, $n_{a}=2, n_{c}=n_{\bar
a}=1$
while 5 electron states cost much energy and are discarded. \\
\begin{figure}[h]
\centerline{\epsfig{figure=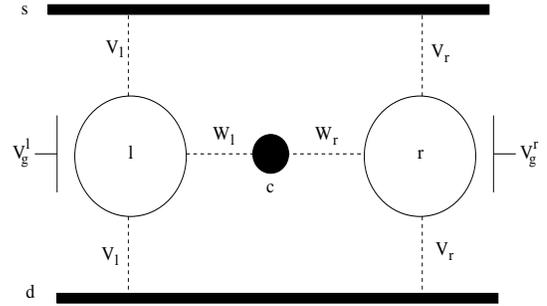,width=70mm,height=40mm,angle=0}}
\caption{Triple quantum dot in parallel geometry.}
\end{figure}
Next, the isolated dot Hamiltonian is diagonalized in the Hilbert
space which is a direct sum of 3 and 4 electron states,
$|\lambda\rangle$ and $|\Lambda\rangle$, using Hubbard operators
$X^{\gamma \gamma}=|\gamma\rangle\langle\gamma|$ ($\gamma=\lambda,
\Lambda$) \cite{Hew}. The four particle states
$|\Lambda\rangle=(|\Lambda_l\rangle, |\Lambda_r\rangle)$ exhaust
the lowest part of the spectrum, an octet consisting of two
singlets $|S_l\rangle,|S_r\rangle$, and two triplets
$|T_l\rangle,|T_r\rangle$. Just above it, there is a charge
transfer exciton $|ex\rangle$. The corresponding energies are,
\begin{eqnarray}
E_{S_{a}} &=& {\epsilon}_c +{\epsilon}_{a}+2{\epsilon}_{\bar a}
+Q_{\bar a} -2W_{a}\beta_{a},\nonumber\\
E_{{T_a}} &=& {\epsilon}_c +{\epsilon}_{a}+2{\epsilon}_{\bar
a}+Q_{\bar a} ,
\nonumber\\
E_{ex} &=& 2{\epsilon}_{l} +2{\epsilon}_{r}+Q_l+Q_r
+2W_{l}\beta_{l}+ 2W_{r}\beta_{r}. \label{En}
\end{eqnarray}
where $\beta_a=W_a/\Delta_a\ll 1$
$(\Delta_a=Q_a+\epsilon_a-\epsilon_c)$. Finally, tunneling
operators in the bare Anderson Hamiltonian are replaced by a
product of number changing Hubbard operators $X^{\lambda \Lambda}$
and a combination
 $c_{k\sigma}=2^{-1/2}(c_{k\sigma s}+c_{k\sigma d})$ of
lead electron operators, ($k=$momentum, $\sigma=$ spin projection
and $s,d$ stand for source and drain).

With these preliminaries, the starting point is a generalized
Anderson Hamiltonian describing the TQD in tunneling contact with
the leads,
\begin{eqnarray}
H_{A}&=& \sum_{k\sigma b=s,d}\epsilon_{kb} c^{\dagger}_{k\sigma
b}c_{k\sigma b} + \sum_{\gamma=\Lambda \lambda} {E}_\gamma
X^{\gamma\gamma} \nonumber\\%
&+&\left(\sum_{\Lambda\lambda} \sum_{k \sigma
a}V^{\Lambda\lambda}_{\sigma a} c^{\dagger}_{k\sigma}X^{\lambda
\Lambda}+ h.c.\right), \label{H}
\end{eqnarray}
with dispersion $\epsilon_{kb}$ of electrons in the leads and
$V^{\lambda\Lambda}_{\sigma a}\equiv V_a\langle\lambda|d_{\sigma
a}|\Lambda\rangle$. The Kondo effect at $T>T_K$ is unraveled by
employing a renormalization group (RG) procedure \cite{Hew,Hald}
in which the energies $E_{\gamma}$  are renormalized as a result
of rescaling high-energy charge excitations. Our attention,
though, is focused on renormalization of $E_{S_{a}},E_{T_{a}}$
(\ref{En}). Since the deep central level $\epsilon_c$ as well as
the tunnel constants are irrelevant variables \cite{KA01,Hald},
the scaling equations are
\begin{equation}
\pi\,dE_\Lambda/d\ln D=\Gamma_\Lambda. \label{DE}
\end{equation}
Here $2D$ is the conduction electron bandwidth, $\Gamma_\Lambda$
are the tunneling strengths,
\begin{equation}
\Gamma_{T_a} = \pi\rho_0 \left(V_{a}^2+ 2V_{\bar{a}}^2\right),\;\;
\Gamma_{S_a} = \alpha_{a}^{2}\Gamma_{T_a},
 \label{GammaR}
\end{equation}
with $\alpha_a=\sqrt{1-2\beta_a^2}$ and $\rho_0$ being the density
of states at $\varepsilon_F$. The scaling invariants for equations
(\ref{DE}),
\begin{equation}
E_{\Lambda}^{\ast }=E_{\Lambda}(D)-\pi^{-1} \Gamma _\Lambda\ln(\pi
D/\Gamma _\Lambda), \label{inv}
\end{equation}
are tuned to satisfy the initial condition
$E_{\Lambda}(D_{0})=E_{\Lambda}^{(0)}$. Equations (\ref{DE})
determine four scaling trajectories $E_\Lambda(D)$ for two singlet
and two triplet states. Note that the level $E_{ex}$ is
irrelevant, but admixture of the bare exciton ($n_{a}=n_{\bar
a}=2$) in the singlet states is crucial for the inequality of
tunneling rates $\Gamma_{T_a}>\Gamma_{S_a}$ (cf.
\cite{KA01,KA02}). As a result, the energies $E_{T_a}(D)$ decrease
with $D$ faster than $E_{S_a}(D)$, so that the trajectories
$E_{T_a}(D, \Gamma_{T_a})$ intersect $E_{S_a}(D, \Gamma_{S_a})$ at
certain points $D^{(a)}=D^{(a)}_{c}$. This level crossing may
occur either before or after reaching the Schrieffer-Wolff (SW)
limit where $E_\Lambda(D)\sim D$ and scaling terminates
\cite{Hald}. Hidden dynamical symmetries affect the Kondo
tunneling most effectively when the scaling trajectories cross
 near the SW boundary
$E_\Lambda(D_c)\sim D_c$. An example of this scenario is shown in
Fig. 2. As a result, various patterns of occasional degeneracy may
arise depending on the initial conditions (\ref{En}), which, in
turn, determine the pertinent $SO(n)$ symmetry of the resulting
spin excitations (see below).

The above Haldane RG procedure brings us to the SW limit
\cite{SW}, where all charge degrees of freedom are quenched. By
properly tuning the SW transformation $e^{iS}$ the effective
Hamiltonian $H=e^{iS}H_{A}e^{-iS}$ is of the $s-d$ type
\cite{Hew}. However, unlike the conventional case \cite{SW} of
doublet spin 1/2  we have here an octet
$\Lambda=\{\Lambda_l,\Lambda_r\}=\{S_l,T_l,S_r,T_r\}$, and the SW
transformation {\it intermixes all these states}. To order
$O(|V|^2)$, then,
\begin{eqnarray}
&&H=\sum_{\Lambda,a}E_{\Lambda_a}X^{\Lambda_a\Lambda_a}+
 \sum_{a}J^T_a {\bf S}_{a}\cdot {\bf s}+ J_{lr}{\hat P}
\sum_a{\bf S}_{a}\cdot {\bf s}\nonumber\\
&&+\sum_{a}J^{ST}_a {\bf M}_{a}\cdot {\bf s} +J_{lr}\sum_{a}{\bf
B}_{a}\cdot {\bf s}+\sum_{k\sigma b}\epsilon_{kb}
c^{\dagger}_{k\sigma b}c_{k\sigma b}. \label{ex}
\end{eqnarray}
The vector operators, ${\bf S}_{a},{\bf M}_{a}, {\bf B}_{a}$ and
the permutation operator ${\hat P}$ manifest the dynamical
symmetry of TQD. Their spherical components are defined via
Hubbard operators connecting different states of the octet:
\begin{eqnarray}
S^{+}_a &=& \sqrt{2}( X^{1_a0_a}+X^{0_a\bar{1}_a}),\ \ S^{-}_a =
(S^{+}_a)^{\dagger}, \nonumber\\
S^z_{a}&=&X^{1_a1_a}-X^{\bar{1}_a\bar{1}_a}, \nonumber\\
 M^{+}_a &=& \sqrt{2}(
X^{1_aS_a}-X^{S_a\bar{1}_a}),\ \ M^{-}_a = (M^{+}_a)^{\dagger},
\nonumber\\
M^z_{a} &=&-(X^{0_aS_a}+X^{S_a0_a}),  \nonumber \\
B^{+}_{a} &=& \sqrt{2}(\alpha_{\bar a} X^{1_aS_{\bar
a}}-\alpha_{a} X^{S_a\bar{1}_{\bar a}}),\ \ B^{-}_{a} =
(B^{+}_{\bar a})^{\dagger},\nonumber\\
B^z_{a} &=& -(\alpha_{\bar a} X^{0_aS_{\bar a}}+\alpha_{a}
X^{S_a0_{\bar a}}). \label{comm}
\end{eqnarray}
Here ${\bf S}_{a}$ are spin 1 operators with projections ${\mu}_a
=1_a,0_a,{\bar{1}_a}$, while ${\bf M}_a$ and ${\bf B}_{a}$ couple
singlet $|S_a\rangle$ with triplet $\langle T_a \mu_{a}|$ and
$\langle T_{\bar a} \mu_{\bar a}|$ respectively. The permutation $
{\hat P} \equiv \sum_{a}(X^{S_a S_{\bar a}}+\sum_{\mu} X^{\mu_a
\mu_{\bar a}})$ commutes with ${\bf S}_l +{\bf S}_r$ and ${\bf
M}_l +{\bf M}_r$, while
${\bf s}=\frac{1}{2}\sum_{kk^{\prime}}\sum_{\sigma\sigma^{%
\prime}} c^{\dagger}_{k'{\sigma}'}\hat{\tau}_{\sigma \sigma'}
c_{k\sigma},$ with Pauli matrices $\hat {\tau}$ is the conduction
electron spin operator. Finally, the (antiferromagnetic) coupling
constants are $ J^T_a=2V_a^2/\Delta_a,\; J^{ST}_a=\alpha_a
J^T_a,\; J_{lr}=V_l V_r \sum_a \Delta_a^{-1}$~
$(\Delta_a=\varepsilon_{F}-\epsilon_a)$.
%%%%%%%%%%%%%%%
\begin{figure}[h]
\centerline{\epsfig{figure=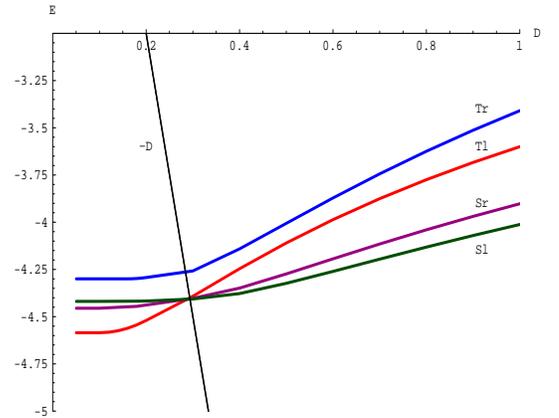,width=70mm,height=55mm,angle=0}}
\caption{Scaling trajectories resulting in an $SO(5)$ symmetry in
the SW regime.}
\end{figure}
After arriving at the $s-d$ Hamiltonian (\ref{ex}) the last stage
implies the solution of Anderson poor-man scaling equations
\cite{Anders} for extracting the corresponding Kondo temperatures.
The discussion below exhausts all possible realizations of $SO(n)$
symmetries arising in TQD.

The most symmetric case is realized when ${\Delta_l}={\Delta_r}$
and ${\Gamma_{T_r}}={\Gamma_{T_l}}$. If all four phase
trajectories $E_{\Lambda}(D)$ intersect at $D=D_c$, the symmetry
of the TQD is $\hat{P}\times SO(4)\times SO(4)$. The operator $\bf
B$ transforms into $\hat{P}{\bf M}$, and the exchange part of $H$
(\ref{ex}) reduces to
\begin{equation}
H_{int}=J^T\sum_{a}(1+{\hat P}){\bf S}_{a}\cdot {\bf s}+
J^{ST}\sum_{a}(1+{\hat P}) {\bf M}_{a}\cdot {\bf s}. \label{sym1}
\end{equation}
The vector operators ${\bf M_a}$ and ${\bf S_a}$ obey the
commutation relations of $o_{4}$ Lie algebra,
\begin{eqnarray}
&&\lbrack
S_{aj},S_{ak}]=ie_{jkm}S_{am},\;\;[M_{aj},M_{ak}]=ie_{jkm}S_{am},
\nonumber\\%
&&[M_{aj},S_{ak}]=ie_{jkm}M_{am}. \label{3.9e}
\end{eqnarray}
(here $j,k,m$ are Cartesian indices). Besides, ${\bf S}_a\cdot
{\bf M}_a=0,$ and the Casimir operator is ${\bf S}_a^{2}+{\bf
M}_a^{2}=3.$
 This justifies the qualification of such
TQD as a {\it double spin rotator.} Scaling equations for $J_T$
and $J_{ST}$ are,
\begin{equation}
\frac{dj_1}{d\ln d} = -2\left[(j_1)^2+(j_2)^2\right],~~
\frac{dj_2}{d\ln d} = -4j_1j_2 , \label{dj}
\end{equation}
with $j_1=\rho_0J^T, j_2=\rho_0J^{ST}, d=\rho_0D$. In the limit of
complete degeneracy the system (\ref{dj}) is reduced to a single
equation, $dj_+/d\ln d = -2(j_+)^2$ for $j_+=j_1+j_2$. Its
solution yields the Kondo temperature
$T_{K0}=\bar{D}\exp(-1/2j_+),$ which is an obvious generalization
of that derived for a QD with $SO(4)$ symmetry and triplet ground
state \cite{Eto,KA01,KA02}. The net spin of the TQD is also $S=1$,
and the residual under-screened spin is $\tilde{S}=1/2$. If the
occasional S/T symmetry is lifted,
$\bar{\delta}=E_S(D_c)-E_T(D_c)>0$, but the TQD still conserves
its permutation symmetry, the Kondo temperature is not universal
anymore, since the scaling of $j_2$ terminates at $D\approx
\bar{\delta}$ (cf. \cite{Eto}). Analytic solution of Eqs.
(\ref{dj}) obtains when $|\bar{\delta}| \gg T_{K0}$, for which
$j_2\approx \alpha j_1$ and $ T_K/T_{K0}\approx
(T_{K0}/\bar{\delta})^{\alpha}. $ The symmetry of TQD in this case
is ${\hat P}\times SO(3)\times SO(3)$.

For the symmetric configurations considered so far, the properties
of TQD are similar to those of DQD, supplemented by the
permutation operation. Much richer are {\it asymmetric}
configurations where ${\Delta_l}\neq{\Delta_r}$,
${\Gamma_{T_l}}\neq{\Gamma_{T_r}}$. When ${\bar E}_{S_l}\approx
{\bar E}_{T_l}\approx {\bar E}_{S_r}< {\bar E}_{T_r}$ (Fig. 2),
the TQD possesses an $SO(5)$ symmetry. The group generators of the
$o_5$ algebra are the "left" vectors ${\bf S}_l, {\bf M}_l$ and
the vector ${\bf B}$ (with $B^+=\sqrt{2}(X^{1_lS_r}-X^{S_r{\bar
1}_l}),$ $B^{-}=(B^{+})^{\dagger},$
$B_z=-(X^{0_lS_r}+X^{S_r0_l})),$ supplemented by the scalar
operator $\hat T =i(X^{S_rS_l}-X^{S_lS_r})$. Thus,
\begin{eqnarray}
&&\lbrack
S_{lj},S_{lk}]=ie_{jkm}S_{lm},\;\;\;\;[M_{lj},M_{lk}]=ie_{jkm}S_{lm},\nonumber\\
&&\lbrack B_{j},S_{lk}]=ie_{jkm}B_{m},\;\;\;\;\;
[B_{j},B_{k}] = ie_{jkm}S_{lm},\nonumber\\
&&\lbrack M_{lj},S_{lk}]=ie_{jkm}M_{lm},\;\;
[M_{lj},B_{k}]=i{\hat T}{\delta}_{jk},\nonumber\\
&&\lbrack
B_{j},{\hat T}] = iM_{lj}, \;\;\;[{\hat T},M_{lj}]=iB_{j},\;%
\;\;[{\hat T},S_{lj}]=0.
\end{eqnarray}
with ${\bf M}_l \cdot {\bf S}_{l}= {\bf B} \cdot {\bf S}_{l}=0,$
${\bf M}_l\cdot {\bf B}=3X^{S_l S_r},$ and Casimir operator ${\bf
S}_l^{2}+{\bf M}_l^{2}+{\bf B}^{2}+T^2=4.$ The exchange
Hamiltonian now reads,
\begin{equation}
{H}_{int} = J_{1l}^T{\bf S}_{l}\cdot {{\bf s}} +J_{1l}^{ST}{\bf
M}_{l}\cdot {{\bf s}}+\alpha_r J_{lr}{\bf B}\cdot {{\bf s}},
\label{int5}
\end{equation}
and the scaling equations are,
\begin{eqnarray}
dj_1/d\ln d &=&
-[j_1^2+j_2^2+j^2_3],\nonumber\\
dj_{2}/d \ln d &=& -2j_1j_2,\;\; dj_3/d\ln d = -2j_1j_3,
 \label{sc5}
\end{eqnarray}
where $j_1=\rho_0 J^{T}_{1l}$, $j_2=\rho_0 J^{ST}_{1l}$ and
$j_{3}=\rho_0\alpha_r J_{lr}$. From Eqs.(\ref{sc5}) the Kondo
temperature is found,
\begin{equation}
T_{K1}=\bar{D}\exp\left\{-\left[j_{1}+\sqrt{j_{2}^2+
j_3^2}\right]^{-1} \right\}. \label{T5}
\end{equation}
Upon increasing $\bar{\delta}_{rl}=E_{S_r}({\bar D})-E_{T_l}({\bar
D})$ the energy $E_{S_r}$ is quenched, and at ${\bar \delta}_r \gg
T_{K1}$ the symmetry reduces to $SO(4)$, with Kondo temperature, $
T_{K}=|{\bar \delta}_r| \exp\{-[j_1(|{\bar\delta}_r|)+ j_2({\bar
\delta}_r)]^{-1}\} $ (cf. \cite{KA02}). On the other hand, upon
decreasing $\bar{\delta}_{r}=E_{T_r}({\bar D})-E_{S_r}({\bar D})$
the symmetry $\hat{P}\times SO(4)\times SO(4)$ is restored at
$\bar{\delta}_{r}<T_{K0}$.

Another "exotic" symmetry, namely $SO(7)$, is realized when the
low-lying multiplet is formed by two triplets $E_{T_{l,r}}$ and
one singlet, say $E_{S_l}$. In this case the $o_7$ algebra is
generated by the six vectors of the type ${\bf S},{\bf M},{\bf B}$
and three scalar operators describing various permutations.
Finally, an $SO(3)$ symmetry occurs when only one triplet state
$E_{Ta}$ (left or right) is relevant, and the $o_3$ algebra is
generated by ${\bf S}_{a}$. The dynamical symmetry of TQD is
thereby exhausted, and summarized by the phase diagram in the
$x,y$ plane with $x=\Gamma_r/\Gamma_l$ and $y=\Delta_l/\Delta_r$
depicted in Fig.3.
\begin{figure}[h]
\centerline{\epsfig{figure=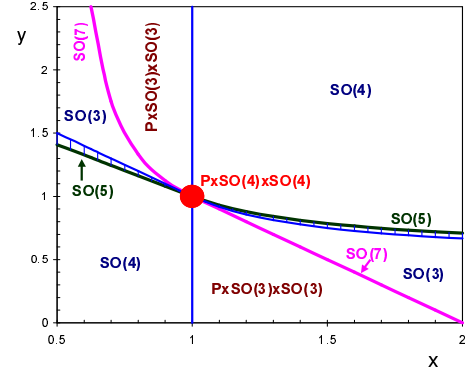,width=60mm,height=49mm,angle=0}}
\caption{Phase diagram of TQD.}
\end{figure}
The central domain of dimension $T_{K0}$ describes the fully
symmetric state (\ref{sym1}), and various regimes of Kondo
tunneling correspond to lines or segments in the $\{x,y\}$ plane.
Vertically hatched domain corresponds to TQD with singlet ground
state where the Kondo effect is absent. Experimental test is
suggested in Fig. 4 which illustrates the evolution of $T_K$, with
$\delta_{rl}\sim y$ for $x=0.96$ and $x=0.7$ corresponding to a
symmetry change from $\hat{P}\times SO(4)\times SO(4)$ to
$\hat{P}\times SO(3)\times SO(3)$ and from $SO(5)$ to $SO(4)$,
respectively.

In similarity with planar QDs or DQDs with $SO(4)$ symmetry
\cite{Misha,Eto,KA01,KA02}, Kondo tunneling may be induced by
external magnetic field $B$ also in the non-magnetic sector of the
phase diagram of Fig. 3 close to the $SO(5)$ line. In this sector
$\bar{\delta}=E_{S_{l,r}}-E_T<0$, and the Kondo effect emerges
when the Zeeman splitting energy $E_z=g\mu_B
B\approx\bar{\delta}$. Due to this compensation $E_{T1}\approx
E_{S_{l,r}}$ and the spin Hamiltonian confined to this subspace
has a form of {\it anisotropic} $SU(2)$ Kondo Hamiltonian
\begin{equation}
\widetilde{H}_{int} = J_\parallel R_zs_z
+J_\perp\left(R^+s^-+R^-s^+\right)/2. \label{int5m}
\end{equation}
Here $J_{\parallel}({\bar D})=J^T_{l}$, $J_\perp({\bar D})=
\sqrt{2[(\alpha_l J^{T}_{l})^2+ (\alpha_r J_{lr})^2]}$. The vector
$\bf R$ is defined as,
\begin{eqnarray}
&&R^+=\sum_a A_aX^{1_lS_a},\;
R^-=(R^+)^{\dagger}, \label{R}\\
&&R_z=[X^{1_l1_l}-\sum_a( A_a^2X^{S_aS_a}+ A_aA_{\bar
a}X^{S_aS_{\bar a}})]/2, \nonumber
\end{eqnarray}
where $A_l=\sqrt{2}\alpha_l J_{\parallel}({\bar D})/J_\perp({\bar
D})$, $A_r=\sqrt{2} \alpha_r J_{lr}/J_\perp({\bar D})$,
$A_l^2+A_r^2=1,$ and $\lbrack R_j,R_k]=ie_{jkm}R_m.$
 The operators (\ref {R})
generate the algebra $o_3$ in the spin subspace $\{S_l, S_r,
T1_l\}$ specified by the Casimir operator $R^2=(3/4)
[X^{1_l1_l}+\sum_a( A_a^2X^{S_aS_a}+ A_aA_{\bar a}X^{S_aS_{\bar
a}})].$ The scaling equations for dimensionless exchange constants
read,
\begin{equation}
dj_{\parallel}/d\ln d = -(j_\perp)^2,\;\;\;\; dj_{\perp}/d\ln d =
-j_{\parallel}j_{\perp},
 \label{sc5m}
\end{equation}
yielding the Kondo temperature,
\begin{equation}
T_{Kz}=\bar{D}\exp\left[-\frac{1}{C}\left(\frac{\pi}{2}-
\arctan\left(\frac{j_{\parallel}}{C}\right)\right)\right],
\label{T5m}
\end{equation}
where $C=\sqrt{(2\alpha_l^2-1)(j_{\parallel})^2+ 2(\alpha_r
j_{lr})^2}$.
%**********
\begin{figure}[h]
\centerline{\epsfig{figure=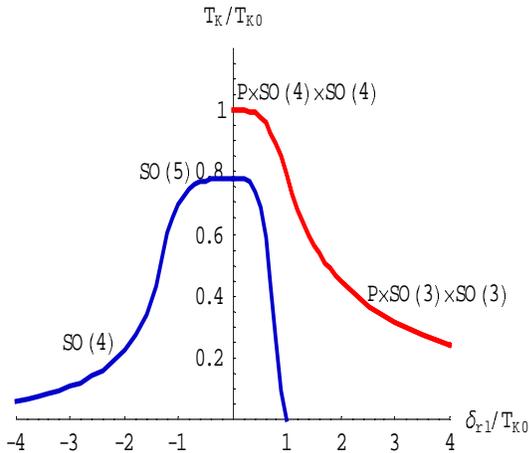,width=70mm,height=60mm,angle=0}}
\caption{Kondo temperature}
\end{figure}
Another type of field induced Kondo effect is realized in
symmetric case of $\bar{\delta}=E_{S_{l,r}}-E_{T_{l,r}}<0$. Now
two components of a triplet, namely $E_{T1_{l,r}}$ cross with two
singlet states, and the symmetry group of the TQD is $SO(4)$. The
$o_4$ algebra is formed by two vectors ${\bf R}$ and $\hat{P}{\bf
R}$ which intermix the states $S_{l,r}$ and $T_{l,r}$. The Kondo
Hamiltonian is also anisotropic. RG procedure similar to
(\ref{sc5m}) yields the Kondo temperature
\begin{equation}
T_{Kz}=\bar{D}\exp\left[-\frac{1}{2C}\left(\frac{\pi}{2}-
\arctan\left(\frac{j^T}{C}\right)\right)\right], \label{T4m}
\end{equation}
where $C=\sqrt{(2\alpha^2-1)(j^T)^2}$.

To conclude, the dynamical $SO(n)$ symmetry of Kondo tunneling
through an evenly occupied TQD is unraveled. It is found that the
Kondo resonance with variable $T_K$ arises due to strong
correlations in a central well, which plays a role of side-coupled
dot for both left and right wells. The hidden dynamical symmetry
manifests itself, firstly in the very existence of the Kondo
effect in QDs with even ${\cal N}$, secondly in non-universal
$T_K$. Its dependence on the ratios $x,y$ of the gate voltages and
tunneling rate may be observed as peculiar conductance curve
$g(x,y)$ at low temperature in specific Coulomb blockade windows,
following the curve $T_K(x,y)$ exemplified in Figs. 3,4. In a
singlet spin state the anisotropic Kondo effect can be induced in
TQD by external magnetic field.
\par The theory is constructed in a single-channel approximation for
lead electrons. In a split gate geometry, more than one tunneling
channel may arise. One may anticipate that the peculiar even
occupation regime of complex QDs then transforms into conventional
odd occupation Kondo regime.
\par This research is partially supported by ISF, BSF and DIP funds.
%%%%%%%%%%%

\end{document}